\documentstyle[12pt]{article}

\newcommand{\be}{\begin{equation}}
\newcommand{\ee}{\end{equation}}
\newcommand{\bea}{\begin{eqnarray}}
\newcommand{\eea}{\end{eqnarray}}

\renewcommand{\theequation}{\arabic{section}.\arabic{equation}}

\begin{document}
\begin{titlepage}

{\flushright{CERN-TH/98-22}}

\vspace{.7in}

\begin{center}
\Large
{\bf Dualities and Hidden Supersymmetry in String Quantum Cosmology}

\vspace{.7in}

\normalsize

\large{James E. Lidsey$^{1a}$ and J. Maharana$^{2b*}$}

\normalsize
\vspace{.3in}

$^1${\em Astronomy Centre and Centre for Theoretical Physics, \\
University of Sussex, BRIGHTON, BN1 9QH, U. K.}

\vspace{.3in}

$^2${\em CERN, Geneva, SWITZERLAND}

\end{center}

\vspace{.7in}

\baselineskip=24pt
\begin{abstract}
\noindent
A supersymmetric approach to string quantum cosmology 
based on the non--compact, global duality symmetries of the 
effective action is developed. An $N=2$ supersymmetric 
action is derived whose bosonic component is 
the Neveu--Schwarz/Neveu--Schwarz sector of the 
$(d+1)$--dimensional 
effective action compactified on a $d$--torus. 
A representation for the  supercharges is 
found and the form of the zero--and one--fermion quantum 
states is determined. 
The purely bosonic component of the wavefunction is unique and 
manifestly invariant under the symmetry of the action. 
The formalism applies 
to a wide class of non--linear sigma--models. 

\end{abstract}

PACS NUMBERS: 04.50.$+$h, 11.25.Mj, 11.30.Pb, 98.80.Hw

\vspace{.1in}
$^a$Electronic mail: jlidsey@astr.cpes.susx.ac.uk \\
$^b$Electronic mail: maharana@nxth04.cern.ch \\
$^*$Permanent Address: Institute of Physics, Bhubaneswar - 751 005, INDIA
 
\end{titlepage}


\section{Introduction}

\setcounter{equation}{0}

\def\theequation{\thesection.\arabic{equation}}

Two approaches to the subject of quantum gravity that have 
received considerable attention in recent years
are quantum cosmology 
 \cite{dewitt} and the superstring theory \cite{gsw}. 
In the canonical quantization of Einstein gravity, 
the classical Hamiltonian, $H=0$, becomes a quantum operator. 
The physical state, $\Psi$, of the universe is then identified as the
eigenstate of this operator with zero eigenvalue, $\hat{H} \Psi 
=0$. This equation decouples into two components, 
$N^i\hat{H}_i \Psi =0$ and 
$N \hat{H}_0 \Psi =0$, where $N^i$ and $N$ 
denote the shift and lapse functions, respectively. 
The first constraint implies 
the invariance of the wavefunction under spatial 
diffeomorphisms and the second is the 
Wheeler--DeWitt equation \cite{dewitt}. In the minisuperspace 
approximation, where inhomogeneous modes are frozen out before 
quantization, this is the only non--trivial constraint 
and can be solved, in principle, by imposing suitable 
boundary conditions \cite{boundary}. 

String theory remains the most 
promising theory for a unification 
of the fundamental interactions. 
It is  now widely thought that the five 
separate theories 
are  non--perturbatively equivalent 
and are related  by `duality' symmetries, 
referred to as S--, T-- and U--duality, respectively 
\cite{duality,ht,witten}. In general, 
these dualities are discrete subgroups of the 
non--compact, global symmetry groups 
of the low--energy effective 
supergravity actions. 
T--duality is a perturbative symmetry in the string 
coupling, but S--duality 
is non--perturbative. 
U--duality interchanges string and sigma--model coupling 
constants and, in this sense, represents a unification of S-- and 
T--dualities. 

Quantum gravitational effects would have played a 
key role in the very early universe  
and this represents one of the few environments 
where predictions of string 
theory may be quantitatively tested. 
A central paradigm of early universe cosmology is that 
of inflation, where the expansion  briefly underwent a very rapid,
acceleration. The above frameworks may be employed to study 
the very early universe
and a question that naturally arises is whether 
they are compatible. 
At present, it is far from clear how such a question 
could be fully addressed. Consequently, a  
more pragmatic  approach is to study how the 
unique features  of string theory, such as its 
duality symmetries, may be 
employed to gain further insight into quantum cosmology, 
and vice--versa.  

In string quantum cosmology, one solves the Wheeler--DeWitt equation
derived from the tree--level string 
effective action \cite{previous,lid,bb}. The
interpretive framework of quantum cosmology may then be employed to
investigate whether string theory leads to realistic cosmologies and,
in particular, whether inflation is probable. 
A quantum cosmological
approach was recently advocated for solving the problem of how
inflation ends in pre-big-bang string cosmology 
\cite{solve,solve1,solve3}. The
well known factor ordering problem is also resolved in this approach
because the symmetries of the action 
imply that the minisuperspace metric 
should be manifestly flat \cite{solve}. Moreover, these symmetries allow
the Wheeler--DeWitt equation to be solved in general for a wide class
of models \cite{solve2}.  For example, in the anisotropic Bianchi type IX
model, the wavefunction becomes increasingly peaked around the
isotropic limit at large spatial volumes \cite{isotropic}.

When restricted to spatially flat, isotropic 
Friedmann--Robertson--Walker (FRW) 
cosmologies, the dilaton--graviton sector of the string effective
action is invariant under an inversion of the scale factor and a shift 
in the dilaton field \cite{sfd}. This 
`scale factor duality' is a subgroup of
T--duality and leads to a supersymmetric extension of the quantum
cosmology, where the classical minisuperspace Hamiltonian 
may be viewed at the quantum level 
as the bosonic component of an $N=2$
supersymmetric Hamiltonian \cite{lid,graham}. This is important because
supersymmetric quantum cosmology may resolve the problems that arise
in the standard approach in constructing a conserved, non--negative
norm from the wavefunction. 
(For a recent review see, e.g., Refs. \cite{moniz,susybook}). 

Thus, string quantum cosmology is well motivated. The purpose of the
present paper is to develop a supersymmetric approach to quantum
cosmology by employing the non--compact, global symmetries of the
string effective action. All ten--dimensional string theories contain
a dilaton, graviton and antisymmetric two--form potential in the
Neveu--Schwarz/Neveu--Schwarz (NS--NS) sector of the
theory. Furthermore, an interesting cosmology is the
spatially flat and homogeneous, Bianchi type I universe admitting $d$
compact Abelian isometries. We therefore consider the NS--NS
sector of the effective 
action compactified on a $d$--torus.  The reduced action
is invariant under a global ${\rm O} (d,d)$ `T--duality', where the
scalar fields parametrize  the coset ${\rm O}(d,d)/[{\rm O}(d) \times
{\rm O}(d)]$.  This leads to an ${\rm O} (d,d)$ invariant
Wheeler--DeWitt equation \cite{solve}.  

The paper is organized as follows. After reviewing the derivation
of the Wheeler--DeWitt equation 
in Section 2, we proceed in Section 3 to derive an
$N=2$ supersymmetric action whose bosonic component is ${\rm O} (d,d)$
invariant. The corresponding super--constraints on the wavefunction
are then derived. These constraints are solved for the zero--fermion
and one--fermion states in Section 4. We conclude 
in Section 5 with a discussion of  how the analysis may be 
extended to a general class of non--linear sigma--models. 

Unless otherwise stated,  units are chosen such that $\hbar =c=1$.

\section{O(d,d) Invariant Wheeler--DeWitt Equation }

\setcounter{equation}{0}

\def\theequation{\thesection.\arabic{equation}}
  
\subsection{Effective Action}

The NS--NS sector of the $(d+1)$--dimensional, tree--level string
effective action is given by \cite{fradkin}
\be
\label{action}
S=\frac{1}{2\lambda_s^{d-1}} \int 
d^{d+1}x\sqrt{|g|} e^{-\Phi} \left[ 
R +\left( \nabla \Phi \right)^2 -\frac{1}{12}
H_{\alpha \beta \gamma}H^{\alpha \beta \gamma} 
+V \right]   ,
\ee
where the Yang--Mills fields are assumed to be trivial,
$\Phi$ is the dilaton field, $V$ 
is an interaction potential, $R$ is the Ricci curvature 
scalar of the space--time with metric ${\cal{G}}$ and 
signature $(-,+,+, \ldots ,+)$, $g \equiv 
{\rm det} {\cal{G}}$, $H_{\alpha\beta\gamma} 
\equiv \partial_{[\alpha} B_{\beta\gamma]}$ is the 
field strength of the antisymmetric two--form 
potential, $B_{\beta\gamma}$,  and $\lambda_s \equiv (\alpha' )^{1/2}$ 
is the fundamental string length scale. 

We assume a spatially closed, flat, 
homogeneous (Bianchi type I) space--time, where the dilaton 
and two--form potential are constant on the surfaces of homogeneity, 
$t = {\rm constant}$. Without loss of generality, 
we may specify ${\cal{G}}_{00} =-1$ and 
${\cal{G}}_{0i}=B_{0i} =0$. Integrating 
over the spatial variables in Eq. (\ref{action}) 
then implies that 
\be
\label{simpleaction}
S=\int d \tau \left[ \bar{\Phi}'^2 +\frac{1}{8} 
{\rm Tr} \left( M' (M^{-1})' \right) +Ve^{-2
\bar{\Phi}} \right]   ,
\ee
where 
\be
\label{shifteddilaton}
\bar{\Phi} \equiv \Phi -\frac{1}{2} 
\ln \left|g \right|
\ee
is the shifted dilaton field, 
\be
\label{dilatontime}
\tau \equiv \int^t dt_1 e^{\bar{\Phi} ( t_1)}
\ee
is the `dilaton' time parameter, 
\be
\label{M}
M \equiv  \left( \begin{array}{cc} G^{-1} & -G^{-1}B \\BG^{-1} & G-BG^{-1}B
\end{array} \right)
\ee
is a symmetric $2d \times 2d$ matrix, 
$G$  is the metric on the 
spatial hypersurfaces, 
a prime denotes differentiation with respect to $\tau$
and we have specified $\lambda_s \equiv 2$ \cite{ven}. 
The dilaton has also been shifted by the constant 
value $\Phi_0 = -\ln (\lambda_s^{-d} \int d^dx)$. 
The matrix $M$ satisfies the 
conditions  
\be
\label{important}
M\eta M =\eta , \qquad M=M^T    ,
\ee
where
\be
\label{eta}
\eta \equiv \left( \begin{array}{cc} 0& I \\ I & 0
\end{array} \right)
\ee
and $I$ is the $d \times d$ unit matrix. It is therefore 
an element of the group ${\rm O} (d,d)$ and its inverse is 
given linearly by $M^{-1} =\eta M \eta$. 

The kinetic sector of action (\ref{simpleaction})
is invariant under a global ${\rm O} (d,d)$ transformation \cite{ven}:
\be
\label{transform}
\tilde{\bar{\Phi}} =\bar{\Phi} , \qquad 
\tilde{M} = \Omega^T M \Omega , \qquad \Omega^T \eta \Omega =\eta  ,
\ee
where $\Omega$ is a constant 
matrix. Since the shifted dilaton field transforms as 
a singlet under the action of Eq. (\ref{transform}), 
this symmetry  is respected when $V$ is an 
arbitrary function of $\bar{\Phi}$. 

The classical Hamiltonian for this cosmological model 
is given by 
\be
\label{bosham}
H_{\rm bos} = \frac{1}{4} \Pi^2_{\bar{\Phi}}
-2{\rm Tr} (M\Pi_M M\Pi_M ) -Ve^{-2\bar{\Phi}}   ,
\ee
where 
\be
\label{momenta}
\Pi_{\bar{\Phi}} =2 \bar{\Phi}' , \qquad 
\Pi_M =-\frac{1}{4} M^{-1} M' M^{-1}
\ee
are the momenta 
conjugate to $\bar{\Phi}$ and $M$, respectively. 
The equations of motion 
for the matrix $M$ can be integrated 
directly to yield the first integral $M\eta M' =C$ \cite{ven}, 
where $C$ is a constant, $2d \times 2d$ matrix satisfying 
the conditions
\be
\label{C} 
C^T=-C, \qquad M\eta C =- C \eta M    .
\ee
The first integral represents a conservation law 
and may also be written in terms of the conjugate momenta 
(\ref{momenta}):
\be
\label{classmom}
M\Pi_M = -\frac{1}{4} C \eta   .
\ee

\subsection{Quantum Cosmology}

The cosmology is quantized  by identifying the
momenta (\ref{momenta}) with the differential operators 
\be
\label{bosop}
\Pi_{\bar{\Phi}} = -i \frac{\delta}{\delta \bar{\Phi}}, 
\qquad   \Pi_{M} = -i\frac{\delta}{\delta M}  .
\ee
Substituting Eq. (\ref{bosop}) into Eq. (\ref{bosham}) then
leads to the Wheeler--DeWitt equation \cite{solve}: 
\be
\label{wdw}
\left[ \frac{\delta^2}{\delta \bar{\Phi}^2}
+8 {\rm Tr} \left( \eta \frac{\delta}{\delta M}
\eta \frac{\delta}{\delta M} \right) +4V e^{-2\bar{\Phi}} 
\right] \Psi (\bar{\Phi}, M ) =0   .
\ee

However, a further constraint 
should also be imposed on the wavefunction because 
the matrix $M$ belongs to the group  
${\rm O}(d,d)$. This results in the conservation law (\ref{classmom}) and 
is analogous to the `rigid rotator'  model
for a particle moving in a 
spherically symmetric potential well. In this model, angular 
momentum is conserved due to 
a global ${\rm O}(3)$ symmetry. When states of 
definite angular momentum are considered, there arises 
a centrifugal barrier term in the effective action and 
we encounter a similar situation
in the model considered above. 
The conservation law (\ref{classmom}) is responsible for the analogue
of the `centrifugal barrier' term when 
the Wheeler-DeWitt equation is solved subject to the 
requirement that the wavefunction satisfies 
the necessary ${\rm O}(d,d)$ invariance properties. 
Thus, condition (\ref{classmom}) should 
apply at the quantum cosmological level and this implies that the 
wavefunction should satisfy the first--order 
constraint  \cite{solve}
\be
\label{quantummom}
i M \frac{\delta \Psi}{\delta M} = \frac{1}{4} C \eta \Psi  .
\ee

In general, the constraint (\ref{quantummom}) 
can not be solved in closed form. On the other hand, it 
does imply that the wavefunction in Eq. (\ref{wdw}) 
can  be separated 
by specifying $\Psi (M, \bar{\Phi} ) = X (M) Y(\bar{\Phi} )$, 
where $X (M)$ and $ Y(\bar{\Phi} )$ are functions of $M$ and 
$\bar{\Phi}$, respectively. 
The Wheeler--DeWitt equation then simplifies 
to an ordinary, differential equation 
in the shifted dilaton field: 
\be
\label{ordinary}
\left[ \frac{d^2}{d \bar{\Phi}^2} + B^2 
 +4 V(\bar{\Phi} )
e^{-2\bar{\Phi}} \right] Y (\bar{\Phi} )=0  ,
\ee
where 
\be
\label{B}
B^2 \equiv \frac{1}{2}
{\rm Tr} \left( C\eta \right)^2
\ee
is a constant and represents the 
`centrifugal barrier' term alluded to earlier.
When $V (\bar{\Phi})$ is constant, 
Eq. (\ref{ordinary}) can be solved 
in full generality in terms of a linear superposition 
of Bessel functions \cite{solve}. 

\section{Supersymmetric String Quantum Cosmology}

\setcounter{equation}{0}

\def\theequation{\thesection.\arabic{equation}}

\subsection{N=2 Supersymmetry}

In this Section we derive 
an $N=2$ supersymmetric Lagrangian whose bosonic sector 
is invariant under global ${\rm O}(d,d)$ transformations. 
In supersymmetric point particle mechanics, the $N=2$ case 
allows an interaction potential to be included. 
Homogeneous $N=2$ supersymmetric quantum cosmologies 
coupled to a single scalar field 
have been studied previously in a different context \cite{gb,bg}. 
It has been further shown that the scale factor duality 
of isotropic FRW string cosmologies is associated with an $N=2$ 
supersymmetry \cite{lid}. We 
extend previous analyses to the class of 
spatially flat, anisotropic 
(Bianchi type I) cosmologies including 
a non--trivial NS--NS two--form 
potential. 

In formulating an $N=2$ supersymmetric action, we consider 
superfields of the generic form
\be
\label{superfield}
X^{\mu} (\tau , \theta , \bar{\theta} ) \equiv x^{\mu} ( \tau ) +
i \bar{\psi}^{\mu} (\tau ) \theta +i \psi^{\mu} (\tau ) 
\bar{\theta} +F^{\mu} (\tau ) 
\theta \bar{\theta}    ,
\ee
where the bosonic functions, $\{ x^{\mu}(\tau) ,F^{\mu} (\tau) \}$,  
and anticommuting complex spinor functions,  
$\{ \psi^{\mu} (\tau) , \bar{\psi}^{\mu} (\tau) \} $, 
are arbitrary functions of the dilaton time (\ref{dilatontime}), 
$\{ \theta , \bar{\theta} \} $ are constant, anticommuting, complex 
spinors and $\mu$ is a parameter labelling the 
degrees of freedom in minisuperspace. 

The generators 
for the supersymmetry are defined by 
\bea
\hat{Q}_1 \equiv  -\frac{\partial}{\partial \bar{\theta}} -i 
\theta \frac{\partial}{\partial \tau} \nonumber 
\\
\hat{Q}_2  \equiv \frac{\partial}{\partial \theta} +i
\bar{\theta} \frac{\partial}{\partial \tau}
\eea
and the supersymmetry transformation rule for the 
superfield (\ref{superfield}) is 
\be
\label{superrule}
\delta X^{\mu} =-i \left( \xi_1 \hat{Q}_1 + 
\xi_2 \hat{Q}_2 \right) X^{\mu}   ,
\ee
where $\xi_i$ are arbitrary parameters that commute 
with the bosonic variables  and anticommute with all fermionic variables. 

We now define the  superfields 
\bea
\label{m}
m_{ij}(\tau , \theta , \bar{\theta} ) \equiv M_{ij} (\tau )
+i\bar{\psi}_{ij} (\tau )  \theta +i 
\psi_{ij} (\tau ) \bar{\theta} +F_{ij} (\tau ) 
\theta \bar{\theta} \\
\label{D}
{\cal{D}} (\tau , \theta , \bar{\theta} ) \equiv 
\sqrt{2}  \bar{\Phi} (\tau) +i \bar{\chi} 
(\tau) \theta +i \chi (\tau ) \bar{\theta} 
+ f( \tau )\theta \bar{\theta}   ,
\eea
where $M_{ij}(\tau)$ is given by Eq. (\ref{M}), 
$\{  \psi_{ij} ,\bar{\psi}_{ij} , \chi ,\bar{\chi} \}$ 
are anticommuting,  
complex spinors and $(i,j) = (1, 2, \ldots , 2d)$. The spatial metric and 
antisymmetric, two--form potential determine the bosonic 
component of the superfield 
(\ref{m}) and the shifted dilaton 
field (\ref{shifteddilaton}) plays the equivalent role in Eq. 
(\ref{D}). 

We then define two further superfields in 
terms of Eqs. (\ref{m}) and (\ref{D}): 
\bea
\label{sigma}
\Sigma (\tau, \theta , \bar{\theta} ) \equiv \frac{1}{8} 
\hat{D}_1 m_{ij} \eta^{jk} \hat{D}_2 
m_{kl} \eta^{li} \\
\label{Y}
Y (\tau , \theta , \bar{\theta} ) \equiv \frac{1}{2} \hat{D}_1 {\cal{D}}
\hat{D}_2 {\cal{D}} -W ( {\cal{D}} )   ,
\eea
where the derivative operators are 
\bea
\hat{D}_1 \equiv - \frac{\partial}{\partial \bar{\theta}}
+i\theta \frac{\partial}{\partial \tau} \\
\hat{D}_2 \equiv \frac{\partial}{\partial \theta}
-i \bar{\theta} \frac{\partial}{\partial \tau} 
\eea
and the potential, $W({\cal{D}})$, is an 
arbitrary function of ${\cal{D}}$.  

The sum of  Eqs. (\ref{sigma}) and 
(\ref{Y}) is viewed as an effective Lagrangian in the action: 
\be
\label{supact}
I_{\rm SUSY}  \equiv \int d\tau \int d\theta d \bar{\theta} 
\left( \Sigma + Y \right)    .
\ee
It may be verified by 
substituting Eqs. (\ref{m}) and (\ref{D}) 
into Eqs. (\ref{sigma}) and (\ref{Y}), 
expanding the potential $W({\cal{D}})$ around 
$\bar{\Phi}$, and collecting coefficients  in
the Grassmann variables  
$\theta$ and $\bar{\theta}$ that the action 
(\ref{supact}) is invariant under 
the supersymmetry transformations (\ref{superrule}) 
up to a total time derivative in the $\theta \bar{\theta}$ 
coefficient. This coefficient is given by $L \equiv 
L_g+L_1$, where 
\bea
\label{lg}
L_g = \frac{1}{8} \left[ 
i \psi_{ij}  \eta^{jk} \bar{\psi}'_{kl} \eta^{li} 
-i \psi_{ij}'\eta^{jk} \bar{\psi}_{kl} 
\eta^{li} +M'_{ij}\eta^{jk}M'_{kl}\eta^{li} \right] \\
\label{l1}
L_1 = \left( \bar{\Phi}' \right)^2 +\frac{i}{2} \left( 
\bar{\chi} \chi' -\bar{\chi}' \chi \right)
+\frac{1}{2} f^2 -\frac{1}{\sqrt{2}} f \partial_{\bar{\Phi}} W
-\frac{1}{4} \left( \partial^2_{\bar{\Phi}} W \right) \left[ 
\bar{\chi}, \chi \right]_-
\eea
and $W=W(\bar{\Phi})$. There is an additional term 
in Eq. (\ref{lg}) of the form $F_{ij}\eta^{jk}F_{kl}\eta^{li}$, but
since there is no potential contribution in 
the action from the matrix (\ref{M}), 
its equation of motion implies that 
we may specify $F_{ij} =0$ without loss 
of generality. On the other hand, the equation of 
motion for the auxiliary field, $f$, is given by 
\be
\label{ffield}
f=\frac{1}{\sqrt{2}} \partial_{\bar{\Phi}} W
\ee
and substituting Eq. (\ref{ffield}) into Eq. (\ref{l1})
eliminates this field from the action. 

Integrating 
over the Grassmann variables 
in Eq. (\ref{supact})  therefore implies that 
the $N=2$ supersymmetric action is given by
\bea
\label{supact1}
I_{\rm SUSY} =\int d\tau \left[ \frac{1}{8} \left( 
i\psi_{ij} \eta^{jk} \bar{\psi}'_{kl} \eta^{li} 
-i\psi'_{ij}\eta^{jk}\bar{\psi}_{kl} \eta^{li} +M'_{ij} 
\eta^{jk}M'_{kl}\eta^{li} \right) \right. \nonumber \\
+  \left. \left( \bar{\Phi}'
\right)^2 +\frac{i}{2} \left( \bar{\chi}\chi' -
\bar{\chi}'\chi \right) -\frac{1}{4} \left( \partial_{\bar{\Phi}}
W \right)^2 -\frac{1}{4} \left( \partial^2_{\bar{\Phi}} W 
\right) \left[ \bar{\chi} , \chi \right]_- \right]   .
\eea
The action (\ref{supact1}) reduces to the bosonic action 
(\ref{simpleaction}) in the limit where the Grassmann 
variables vanish if we identify the potential $W$:
\be
\label{identify}
\left( \partial_{\bar{\Phi}} W \right)^2 =-4 
V(\bar{\Phi} ) e^{-2\bar{\Phi}}   .
\ee
Thus, a necessary condition for an $N=2$  supersymmetric 
extension of the effective action (\ref{simpleaction}) is that 
the interaction potential, $V$, must be  semi--negative 
definite.

The classical momenta conjugate to the bosonic and 
fermionic degrees of freedom in action (\ref{supact1}) 
are
\bea
\label{mom1}
\Pi_{M_{mn}} = \frac{\partial L}{\partial M'_{mn}} = 
\frac{1}{4} \eta^{nk} M'_{kl} \eta^{lm} \nonumber \\
K_{mn} = \frac{\partial L}{\partial \psi'_{mn}} 
=-\frac{i}{8} \eta^{nk} \bar{\psi}_{kl} \eta^{lm} 
\nonumber \\
\bar{K}_{mn} =\frac{\partial L}{\partial \bar{\psi}'_{mn}}
=-\frac{i}{8} \psi_{ij} \eta^{jm} \eta^{ni} \nonumber \\
\Pi_{\bar{\Phi}} = \frac{\partial L}{\partial \bar{\Phi}'} = 2 
\bar{\Phi}' \nonumber \\
\Pi_{\chi} = \frac{\partial L}{\partial \chi'} = -\frac{i}{2} 
\bar{\chi} \nonumber \\
\Pi_{\bar{\chi}} = \frac{\partial L}{\partial \bar{\chi}'} = 
-\frac{i}{2} \chi  ,
\eea
respectively, where the negative sign appears in the expressions for 
$\bar{\psi}_{ij}$ and $\chi$ because the left 
derivative of the Grassmann variables is taken. 
The classical Hamiltonian 
derived from the action (\ref{supact1}) is given by 
\be
\label{hamg}
H = M'_{ij}\eta^{jk}\Pi_{M_{kl}}\eta^{li}
+\psi'_{ij}\eta^{jk}K_{kl}\eta^{li} 
+\bar{\psi}'_{ij}\eta^{jk}\bar{K}_{kl}\eta^{li} + \bar{\Phi}' 
\Pi_{\bar{\Phi}} + \chi\Pi_{\chi} + \bar{\chi} \Pi_{\bar{\chi}} -L
\ee
and substituting Eqs. (\ref{mom1}) into Eq. (\ref{hamg}) implies 
that it takes the form
\be
\label{classham}
H= 2\Pi_{ij}\eta^{jk}\Pi_{kl}\eta^{li} +
\frac{1}{4}\Pi^2_{\bar{\Phi}}+
\frac{1}{4}\left( \partial_{\bar{\Phi}} W 
\right)^2 +\frac{1}{4} \left( 
\partial^2_{\bar{\Phi}} W \right)
\left[ \bar{\chi} ,\chi \right]_-   ,
\ee
where the anticommuting property of the 
Grassmann variables has been employed and $\Pi_{ij} 
\equiv \Pi_{M_{ij}}$. The 
bosonic component 
of the Hamiltonian (\ref{classham}) corresponds to 
Eq. (\ref{bosham}). 
The first term in this expression describes the Hamiltonian 
for the matrix $M_{ij}$. 
The fermions do not appear in this component of the Hamiltonian, 
as is always the case in supersymmetric quantum 
mechanics when the fermions are free.

\subsection{Quantum Constraints}

The model is quantized by assuming 
the standard operator realizations for the bosonic variables
in Eq. (\ref{bosop}) 
and further imposing the spinor algebra 
\bea
\label{gral}
\left[ \psi_{ij} ,\psi_{kl} \right]_+ = 
\left[ \bar{\psi}_{ij},\bar{\psi}_{kl} \right]_+ =0, 
\qquad \left[ \psi_{ij} , \bar{\psi}_{kl} \right]_+ =
\eta_{ik}\eta_{jl} \nonumber \\
\left[ \chi , \chi \right]_+ =\left[ \bar{\chi},\bar{\chi} 
\right]_+ =0 , \qquad \left[ \chi ,\bar{\chi} \right]_+ =1
\nonumber \\
\left[ \bar{\chi} , \bar{\psi}_{ij} 
\right]_+ =\left[ \chi ,\psi_{ij} \right]_+ = 
\left[ \chi ,\bar{\psi}_{ij} \right]_+ = \left[ 
\bar{\chi}, \psi_{ij} \right]_+ =0   .
\eea
A  representation satisfying Eq. (\ref{gral}) 
is given in terms of the set of Grassmann variables $\{
\zeta_{ij} ,\beta \}$: 
\bea
\psi_{kl}=\eta_{kp}\frac{\partial}{\partial \zeta_{pr}} 
\eta_{rl} , \qquad \bar{\psi}_{ij} =\zeta_{ij} \nonumber \\
\chi =\frac{\partial}{\partial \beta} , \qquad \bar{\chi} =
\beta   .
\eea

Notice that we have imposed 
$[\psi_{ij} ,{\bar \psi}_{kl}]_+ =\eta_{ik}\eta_
{jl}$ in Eq. (\ref{gral}). 
The canonically conjugate momenta for $\psi_{ij}$ and $\bar{\psi}_{kl}$ 
are given by Eq. (\ref{mom1}) 
and, if we had employed the canonical anticommutation relations
between $\psi_{ij}$ and $K_{mn}$ 
and between $\bar{\psi}_{ij}$ and $\bar{K}_{mn}$, 
we would not have obtained 
the anticommutation relation presented in Eq. (\ref{gral}). 
We have
implicitly employed $[\psi_{ij} ,K_{kl}]_+ =4
\eta_{ik}\eta_{jl}$ and $[\bar{\psi}_{ij}, \bar{K}_{kl}]_+ =4
\eta_{ik} \eta_{jl}$ rather
than `one' on the right hand side. 
This is due to the fact that there is a factor of $1/8$ in
the kinetic energy terms of $\psi_{ij}$ and 
$\bar{\psi}_{kl}$ rather than the conventional factor of $1/2$ 
that appears in the Dirac Langrangian. (For example, 
the $\chi$ and
$\bar \chi$ pieces in Eq. (\ref{supact1})
have the standard factor). We adopted the
standard anticommutation relation between $\psi_{ij}$ 
and $\bar{\psi}_{kl}$ because the 
calculations can then be performed in a straightforward 
manner without keeping track of additional 
numerical constants. If
the anticommutation relation in Eq. (\ref{gral}) had 
been modified with another
numerical constant, our expressions for the supercharges 
$Q$ and $\bar Q$ 
given below would also have been modified. However, 
an identical expression 
for the Hamiltonian would have been obtained from the 
anticommutator of $Q$ and $\bar Q$.

We now define the supercharges
\bea
\label{sup1}
Q \equiv 2\Pi_{ij}\eta^{jk}\psi_{kl}\eta^{li}
+\frac{1}{\sqrt{2}}\left( 
\Pi_{\bar{\Phi}} +i\partial_{\bar{\Phi}} W 
\right) \chi \\
\label{sup2}
\bar{Q} \equiv 2\Pi_{mn}\eta^{nr}\bar{\psi}_{rp}
\eta^{pm}+\frac{1}{\sqrt{2}}\left( 
\Pi_{\bar{\Phi}} - i \partial_{\bar{\Phi}} W \right) 
\bar{\chi}  ,
\eea
where $Q$ is a non--Hermitian, linear operator 
and $\bar{Q}$ is its adjoint. Substituting 
Eqs. (\ref{sup1}) and (\ref{sup2}) into Eq. (\ref{classham}), 
and employing the anticommutation relations (\ref{gral}), implies that 
the Hamiltonian operator may be written as
\be
\label{alg}
2H=\left[ Q , \bar{Q} \right]_+ ,\qquad Q^2=\bar{Q}^2 =0  ,
\ee
where $[H,Q]_-=[H,\bar{Q}]_- =0$. Thus, there exists an $N=2$ 
supersymmetry in the quantum 
cosmology \cite{graham,witten1}. 
This may be viewed as a direct extension of the 
${\rm O}(d,d)$ `T--duality' of the toroidally 
compactified NS--NS action (\ref{action}). 

Finally, supersymmetry implies that 
the wavefunction of the universe is annihilated 
by the supercharges, $Q\Psi =\bar{Q} \Psi =0$. These 
reduce to a set of first--order differential equations and 
we derive solutions to these constraints in the following Section.   

We conclude this Section by remarking that factor 
ordering problems in supersymmetric quantum mechanics 
in curved space have been addressed previously 
\cite{curve}. In general, 
four fermion terms with the curvature tensor appear, but we 
have not considered such problems here because we assumed 
a toroidal compactification and this implies that the curvature is zero. 
Moreover, as discussed by Gasperini {\em et al.} \cite{solve}, 
the operator 
ordering issue is settled in the standard procedure by demanding the 
${\rm O}(d,d)$ invariance of the Hamiltonian. 	

\section{Quantum States}

\setcounter{equation}{0}

\def\theequation{\thesection.\arabic{equation}}

\subsection{Zero--fermion State}

The quantum constraints are solved 
by defining the conserved `fermion number': 
\be
F \equiv \bar{\psi}_{ij} \eta^{jk} \psi_{kl}
\eta^{li} + \bar{\chi}\chi  ,
\ee
where 
\be
[H, F]_- =0, \qquad [Q,F]_- =Q, \qquad [\bar{Q} ,F]_- =-\bar{Q}  .
\ee
This implies that states with a fixed fermion number may be 
individually considered. 
The fermion vacuum, $|0\rangle$, is defined:  
\be
\label{bosstate}
\psi_{ij} | 0 \rangle =\chi |0\rangle =0 \qquad \forall \qquad
i,j .
\ee
The state with zero fermion number, $| \psi_0 \rangle$, 
is defined as $| \psi_0 \rangle \equiv h(M_{ij} ,\bar{\Phi}) 
| 0 \rangle$, where $h$ is an arbitrary function. This state 
is a function of the bosonic 
degrees of freedom only and is automatically annihilated by the supercharge 
$Q$. It is annihilated by its adjoint, $\bar{Q}$, if the conditions
\be
\label{bosform}
\frac{\delta h}{\delta M_{ij}} =0 , 
\qquad \left[ 
\frac{\partial}{\partial \bar{\Phi}} +  
\frac{dW}{d \bar{\Phi}} \right] h =0
\ee
are simultaneously satisfied. 
Modulo a constant of proportionality, the general 
solution to Eq. (\ref{bosform}) is 
\be
\label{zerof}
| \psi_0 \rangle = e^{-W (\bar{\Phi} )} | 0\rangle   .
\ee
This solution is {\em uniquely} 
determined by the potential (\ref{identify}) and is a function 
of the shifted dilaton field only. It is therefore manifestly 
invariant under the global ${\rm O}(d,d)$ symmetry 
(\ref{transform}) of the 
bosonic action (\ref{simpleaction}). 

It is interesting to relate Eq. (\ref{zerof}) 
to solutions   of the 
Euclidean Hamilton--Jacobi equation. This is given by 
\be
 \label{ehj}
\left( \frac{\delta I}{\delta \bar{\Phi}} \right)^2 
-8 M_{ij} \frac{\delta I}{\delta M_{jk}} 
M_{kl} \frac{\delta I}{\delta M_{li}}  =-4 
V(\bar{\Phi}) e^{-2\bar{\Phi}}  ,
\ee
where $I=I(M_{ij} ,\bar{\Phi} )$ is a Euclidean 
action of the classical theory. The Euclidean analogue of the 
momentum constraint (\ref{classmom}) implies that separable solutions 
to Eq. (\ref{ehj}) can be found by substituting in the ansatz 
$I(M_{ij} ,\bar{\Phi}) =F(M_{ij})+G(\bar{\Phi})$, 
where $F$ and $G$ are functions of 
$M_{ij}$ and $\bar{\Phi}$, respectively. Although 
a closed expression for $F(M_{ij})$ 
can not be determined in general, 
the form of $G(\bar{\Phi})$ can be found in terms 
of quadratures for an arbitrary dilaton 
potential $V(\bar{\Phi})$. It follows that
\be
\label{G}
G(\bar{\Phi}) = \int^{\bar{\Phi}} 
d\bar{\Phi}_1 \left[ B^2 -4V(\bar{\Phi}_1) e^{-2
\bar{\Phi}_1} \right]^{1/2}   ,
\ee
where $B^2$ is defined in Eq. (\ref{B}).   

Comparison with 
Eq. (\ref{identify}) therefore implies that 
the exponent,  $W(\bar{\Phi})$, in 
Eq. (\ref{zerof}) may be interpreted as a Euclidean 
action of the cosmology when $B^2=0$ and $I$ is independent 
of $M_{ij}$. This 
corresponds to the case where the momenta conjugate to 
$M_{ij}$ vanish. The bosonic wavefunction 
(\ref{zerof}) represents an approximate WKB Euclidean  solution to 
the Wheeler--DeWitt equation (\ref{wdw}) when $B^2=0$ but it is 
an exact state in the  
supersymmetric quantization, and moreover, is 
uniquely selected by the symmetry. 

To lowest--order, the potential in the (super--) string effective action 
(\ref{action}) is a cosmological 
constant  determined by the dimensionality of space--time, 
$V \equiv \Lambda = 2(d-9)/3\alpha'$. In this case the cosmological 
constant is 
related to the central charge deficit. We can put forward an argument that
supersymmetric quantum cosmology 
preserving the ${\rm O}(d,d)$ T-duality implies an 
upper bound, $d \le 9$, on the number of spatial dimensions 
in the universe. If one
interprets $P=|\Psi |^2$ as an unrenormalized probability 
density, then $P \propto \exp( - 4 | \Lambda |^{1/2} \Omega_V g_s^{-2})$, 
where $\Omega_V$ is the proper spatial volume and $g_s \equiv e^{\Phi /2}$ 
is the string coupling. 
It is interesting that the probability density 
is peaked in the strong--coupling regime. 
Furthermore, for fixed coupling, the value of 
$P$ increases as $\Lambda \rightarrow 0^-$. In this sense, 
therefore, smaller values of $| \Lambda |$ are favoured.

\subsection{One--fermion State}

We now proceed to find the form of the one--fermion state, $|
\psi_1 \rangle$. A general ansatz  for this component of the 
wavefunction is given in terms of the fermion vacuum by 
\be
\label{genant}
| \psi_1 \rangle = \left( f_{ij} \eta^{jk} \bar{\psi}_{kl} 
\eta^{li} +f_{\chi} \bar{\chi} \right) e^{-W} |0\rangle   ,
\ee
where $f_{ij}$ and 
$f_{\chi}$ are arbitrary functions of the bosonic 
variables over the configuration space. 
Operating on this state with $\bar{Q}$ 
and employing Eq. (\ref{gral}) 
implies that it is annihilated 
if $f_{ij} \equiv 2  \Pi_{ij} f$ and $f_{\chi} \equiv 
(\Pi_{\bar{\Phi}} -i \partial_{\bar{\Phi}}W)f/\sqrt{2}$, 
respectively, where $f=f(M_{ij}, \bar{\Phi})$ is an arbitrary 
function. Substituting these definitions into Eq. (\ref{genant})
then implies that 
\be
\label{onef}
| \psi_1 \rangle \equiv \bar{Q} f e^{-W 
(\bar{\Phi} )}  | 0\rangle  .
\ee
Operating on Eq. (\ref{onef}) with $Q$
and employing Eq. (\ref{alg}) implies that 
\be
\label{other}
Q| \psi_1 \rangle =2H fe^{-W}|0\rangle =0  .
\ee
By employing Eqs. (\ref{gral}) and (\ref{bosstate}), 
it follows that $[\bar{\chi} ,\chi ]_- | 0 \rangle =-|0 
\rangle$. Eq. (\ref{other}) is therefore 
satisfied if $f$ is a solution to the differential 
equation: 
\be
\label{fwdw}
\left[ 2\frac{\delta}{\delta M_{ij}} \eta_{jk} \frac{\delta}{\delta M_{kl}}
\eta_{li} +\frac{1}{4} \frac{\partial^2}{\partial \bar{\Phi}^2}
-\frac{1}{2} \frac{dW}{d\bar{\Phi}} \frac{\partial}{\partial
\bar{\Phi}} \right] f(M_{ij} , \bar{\Phi}) =0   .
\ee

For a separable solution, $f \equiv X(M_{ij} ) 
Y(\bar{\Phi })$, Eq. (\ref{fwdw}) implies that 
\be
\label{fwdwsep}
\left[ \frac{d^2}{d\bar{\Phi}^2} -2 \frac{dW}{d\bar{\Phi}} 
\frac{d}{d\bar{\Phi}} +c^2 \right] Y(\bar{\Phi}) =0
\ee
and 
\be
\label{fMsep}
\left[ \frac{\delta}{\delta M_{ij}} \eta_{jk} 
\frac{\delta}{\delta M_{kl}}
\eta_{li} -\frac{c^2}{8} \right] X(M_{ij}) =0   ,
\ee
where $c$ is a separation constant. In the special case where 
$c=0$, Eq. (\ref{fwdwsep}) may be solved exactly:
\be
Y = \int^{\bar{\Phi}} d\bar{\Phi}_1 e^{2W(\bar{\Phi}_1 )}
\ee
for an arbitrary potential $W(\bar{\Phi} )$. 

In general, Eq. (\ref{fwdwsep}) can not 
be solved in closed form for arbitrary 
$W(\bar{\Phi})$ and $c$. 
However, for a constant dilaton potential, $V=
\Lambda$, we may define the new variable
\be
z \equiv \frac{1}{4|\Lambda |^{1/2}} e^{\bar{\Phi}}
\ee
This implies that Eq. (\ref{fwdwsep}) takes the form
\be
\label{Gfunction}
\left[ z^2 \frac{d^2}{dz^2} +(z-1) \frac{d}{dz} + {c^2} 
\right] Y=0
\ee
and Eq. (\ref{Gfunction}) can be solved in the limit $z \gg 1$. 
A detailed study of this equation  
is beyond the scope of the present work, however. 

\subsection{An SL(2,R) Subgroup}

Thus far, we have constructed 
the ${\rm O}(d,d)$ invariant supersymmetric
Hamiltonian and derived the 
corresponding super--constraint equations 
in the general setting. Moreover, we have 
obtained the state in the zero fermion sector that is 
annihilated  by the supercharges. 
This is uniquely determined by the
potential given in Eq. (\ref{identify}).  
It is rather difficult to proceed further 
with the general form of the M--matrix (\ref{M}), however. 
In view of this, we
now consider a more simple scenario, 
where we deal with an ${\rm SL}(2,R)$
matrix corresponding to a subgroup of ${\rm O}(d,d)$. 

The 
${\rm O}(d,d)$ group may be written as a product 
of ${\rm SL}(2,R)$ subgroups when the components of the 
metric and two--form potential are identified in an appropriate 
fashion. For example, there are three 
${\rm SL}(2,R)$ 
subgroups for $d=6$ \cite{km}. Here we consider just one of them to 
illustrate our
main points. It is then possible to make further progress 
in finding the solutions to the Wheeler--DeWitt equation. It is
worth remarking at this stage 
that the group  ${\rm SL}(2,R)$ also 
appears within the context of the S-duality
group and thus, as we shall discuss later, this analysis is useful in the
context of type IIB string cosmology. The ${\rm SL}(2,R)$ 
subgroup we consider here is part of the T--duality group. 

It is well known that an ${\rm SL}(2,R)$ matrix can be introduced
with unit determinant: 
\be 
\label{TT}
T \equiv\left( \begin{array}{cc} e^{q\Delta}+e^{-q\Delta}{\cal P}^2 &
e^{-q\Delta}{\cal P} \\ e^{-q\Delta}{\cal P} & e^{-q\Delta} \end{array}
\right) ,
\ee
where $q$ is a constant. The two scalar moduli  fields,
$\Delta$
and ${\cal P}$, parametrise the coset 
${{\rm SL}(2,R)/{{\rm SO}(2)}}$. Let us
consider another ${\rm SL}(2,R)$ matrix
\be 
V= \left( \begin{array}{cc} e^{{1\over 2}q\Delta} & 0 \\
e^{-{1\over 2}q\Delta}{\cal P} & e^{-{1\over 2}q\Delta} \end{array} 
\right)   .
\ee 
The matrix $V$ is also of unit determinant and can be thought of
intuitively as the `vielbein' of the ${\rm SL}(2,R)$ metric, 
whereas $T$ is like the
metric itself,  since $T=V^TV$. Notice that under a global
${\rm SL}(2,R)$ transformation, ${\cal G}$, 
and a local ${\rm SO}(2)$ transformation,
$O$, $V \rightarrow OV{\cal G}$. Thus, for a given ${\cal G}$, an $O$
can always be chosen that preserves the form of 
$V$. Thus, the symmetric
matrix, $T$, transforms as $T \rightarrow {\cal G}^T T{\cal G}$.

The action (\ref{simpleaction}) 
can be written in an ${\rm SL}(2,R)$ invariant form by 
replacing the kinetic term for the M--matrix by 
${\rm Tr} [T'(T^{-1})']/4$. The inverse of $T$ may be 
written linearly as $ T^{-1} = -JTJ$, where 
\be
\label{J}
J= \left( \begin{array}{cc} 0 & 1 \\
-1 & 0
\end{array} \right)  ,\qquad J^2 =-I   .
\ee
The classical  
equations of motion for the moduli fields then imply that momentum conjugate 
to ${\cal{P}}$, $\Pi_{\cal{P}} = -
{\cal{P}}'\exp (-2q \Delta )$, is conserved. Thus, 
the quantum constraint  (\ref{quantummom}) takes the simple 
form $i \partial \Psi   /\partial {\cal{P}} =L_{\cal{P}} \Psi$, 
where $L_{\cal{P}}$ is an arbitrary constant. For 
separable solutions, $\Psi \equiv X(T)Y(\bar{\Phi})$, 
the $X(T)$ component of the wavefunction can then be evaluated  
by separating the Wheeler--DeWitt equation (\ref{wdw}). 
The component form of Eq. (\ref{wdw}) may be derived by identifying 
$M_{ij}$ and $\eta_{kl}$ with Eqs. (\ref{TT}) and (\ref{J}), respectively. 
Alternatively, for this model it may be derived directly at the 
level of the Lagrangian. It follows that 
\be
\label{beseqn}
\left[ \frac{1}{2q^2} \frac{\partial^2}{\partial \Delta^2} +
\frac{1}{2} e^{2q\Delta} \frac{\partial^2}{\partial {\cal{P}}^2}
+ B^2 \right] X (T) =0
\ee
and the general solution to this equation that is consistent 
with the first--order momentum constraint (\ref{quantummom}) 
is given by
\be
X =e^{-iL_{\cal{P}} {\cal{P}}}
Z_{\sqrt{2} B} \left( L_{\cal{P}} e^{q \Delta} \right)   ,
\ee
where $Z_{\sqrt{2} B}$ is a linear 
combination of modified Bessel 
functions of order $\sqrt{2}B$. 

The supersymmetric extension may be performed for this 
model as outlined in Section 3 
by defining a superfield, $
N_{ij} \equiv 
T_{ij} +i\bar{A}_{ij} \theta +i A_{ij} \bar{\theta} +C_{ij} 
\theta \bar{\theta}$, analogous to Eq. (\ref{m}). 
Similar conclusions therefore apply and, in particular, 
the structure of Eq. (\ref{fMsep}) for the one--fermion state is
formally equivalent to that of Eq. (\ref{beseqn}). Thus, the solutions 
are again given in terms of modified Bessel functions. 

\section{Conclusion and Discussion}

\setcounter{equation}{0}

\def\theequation{\thesection.\arabic{equation}}

In this paper, we have derived an $N=2$ supersymmetric  quantum 
cosmology from the toroidally compactified string effective action. 
The supersymmetric Hamiltonian operator 
reduces to the ${\rm O}(d,d)$ invariant Hamiltonian in the classical 
limit. The existence of the supersymmetry imposes strong constraints on the 
wavefunction of the universe
and implies that it should by annihilated 
by the supercharges. These are first--order constraints and 
correspond to the Dirac--type 
square root of the Wheeler--DeWitt equation (\ref{wdw}). 
Solutions to these constraints were found for the 
zero-- and one--fermion states. The general 
form of the bosonic component of the wavefunction 
was determined and found to be 
invariant under the non--compact, global 
${\rm O}(d,d)$ symmetry (T--duality) of the classical action. 

The 
supersymmetric approach we have employed may be 
applied to a wide  class of non--linear sigma--models.   
The generalized sigma--model action in the `Einstein' frame is 
given by \cite{gibbons}
\be
\label{actionsigma}
S=\int d^4 x \sqrt{-g} \left[ R  - \frac{1}{2}
\gamma^{ij} (\phi) \nabla^{\mu}\phi_i
\nabla_{\mu} \phi_j -2 \Lambda  \right]   ,
\ee 
where the scalar fields
$ \{ \phi_i \}$ 
may be viewed as coordinates on a target space with metric
$\gamma_{ij}$, 
$\Lambda$ is a cosmological constant and 
units are chosen such that $16\pi G \equiv 1$. 
We assume that the 
target space is a non--compact, Riemannian, symmetric space $G/H$, where 
$G$ is a non--compact Lie group with a maximal compact subgroup $H$
\cite{He}. Eq. (\ref{actionsigma}) 
represents the bosonic sector of many four--dimensional 
supergravity theories when the gauge fields are trivial
and $\Lambda =0$, 
including type II 
theories compactified to four dimensions and all those with $N\ge 
4$ supersymmetry \cite{ht}. The metric $\gamma_{ij}$ may be written as 
\be
\label{tsmetric}
dl^2 =\gamma_{ij}  d\phi^i d\phi^j \equiv {\rm Tr} 
\left( dN dN^{-1} \right) ,
\ee
where the matrix $N$ is an element 
of the group $G$. Substituting Eq. (\ref{tsmetric}) into 
Eq. (\ref{actionsigma}) then implies that 
\be
\label{actiontrace}
S=\int d^4 x \sqrt{-g} \left[ R - \frac{1}{2}
{\rm Tr} \left( \nabla N \nabla 
N^{-1} \right) -2 \Lambda \right]   .
\ee

We now consider the  spatially homogeneous and isotropic
FRW universes with a line element given by 
$ds^2 =-dt^2 +e^{2\alpha (t)} d\Omega_k^2$, 
where $d\Omega_k^2$ is the line element on the three--space 
with constant curvature, $k=\{ 
-1, 0, +1 \}$ for negatively
curved, flat and positively curved models, respectively, 
and $a(t) \equiv e^{\alpha (t)}$ 
is the scale factor of the universe. 
Integrating over the spatial 
variables in action (\ref{actiontrace}) implies that 
\be
\label{actionT}
S=\int dT \left[ -6 (\partial_T \alpha )^2 + \frac{1}{2} {\rm Tr} 
\left( \partial_T N  \partial_T N^{-1} 
\right) - V_{\rm eff} (\alpha ) \right]   ,
\ee
where $\phi_i =\phi_i (t)$, 
$\partial_T$ denotes differentiation with 
respect to the rescaled time variable
\be
\label{T}
T \equiv \int^t dt_1 e^{-3\alpha (t_1)}
\ee
and  
\be
\label{eff}
V_{\rm eff} (\alpha )  \equiv 2\Lambda e^{6\alpha} 
 -6ke^{4\alpha}   .
\ee

Eqs. (\ref{simpleaction}) and 
(\ref{actionT})  are formally very similar and, 
for constant $\Lambda$, the latter 
is invariant under the global action of the group $G$: 
\be
\label{Gsymmetry}
\bar{\alpha} = \alpha , \qquad \bar{N} = \Omega^T N \Omega ,
\ee
where $\Omega \in G$ is a constant matrix.  Since the 
logarithm of the scale 
factor transforms as a singlet, its role is equivalent to that of
the shifted dilaton field (\ref{shifteddilaton}) and 
 Eq. (\ref{eff}) then represents a $G$--invariant 
effective potential for the scale factor. This implies that 
supersymmetric quantum  cosmologies may be
derived from the non--linear sigma--model (\ref{actionT}) 
when the matrix $N$ satisfies appropriate linearity conditions 
analogous to Eq. (\ref{important}). In the ${\rm O}(d,d)$ model, 
the metric (\ref{eta}) satisfies $\eta^2 =I$ and, 
together with Eq. (\ref{important}), this implies 
the important relation
$M^{-1} =\eta M\eta$. Consequently, the  target space metric may 
be written uniquely in terms 
of $M$ and $\eta$. 
Thus, the analysis of Section 3 applies 
directly to {\em all} non--linear sigma--models where $N$ 
is symmetric and its inverse 
is given linearly by 
\be
\label{necessary}
N^{-1} = \pm  
\theta N \theta, \qquad \theta^2 =\pm I, \qquad \theta \in G   .
\ee

For example, it has recently been conjectured that the five 
string theories have a common origin in a new quantum theory, 
referred to as M--theory \cite{witten}. 
The low--energy limit of this theory is 
$N=1$, eleven--dimensional supergravity and this leads to 
$N=8$ supergravity after 
toroidal compactification to four dimensions \cite{cj}. 
The bosonic sector 
of this theory admits 28 Abelian vector gauge fields and 70 scalar 
fields 
that take values in the homogeneous coset space ${\rm E}_{7(7)}/
[{\rm SU}(8)/{\rm Z}_2]$. The discrete 
subgroup ${\rm E}_{7} (Z)$ is the conjectured U--duality 
of type II string theory compactified on a six--torus 
\cite{ht}. 
When the gauge fields are frozen, the effective action 
has the form given by Eq. (\ref{actiontrace}) with $\Lambda =0$. 
In this case, 
$N$ is a symmetric matrix in ${\rm E}_{7(7)}$ that may 
be viewed as a positive metric in the internal 
space corresponding to the 56--dimensional fundamental 
representation of ${\rm E}_{7(7)}$ \cite{cj}. The symplectic 
invariant of ${\rm E}_{7(7)}$ is 
\be
\theta =
\left( \begin{array}{cc} 0 & -I \\
I & 0
\end{array} \right)  , \qquad \theta^2 = 
\left( \begin{array}{cc} -I & 0 \\
0 & -I 
\end{array} \right)   ,
\ee
where $I$ is the $28\times 28$ unit matrix and the 
linearity  condition (\ref{necessary}) is therefore 
satisfied for this model.

In view of various evidences that M-theory provides intricate relations
among the five string theories, there have been attempts to explore the
cosmological implications of M--theory and one is naturally led
to study
cosmological scenarios in type IIA and IIB theories with the effects
of the Ramond-Ramond sector included \cite{jnansref}. 
It would be interesting to investigate
quantum cosmologies for M-theory and string theories along the lines
followed in this work.

When $d=3$, Eq. (\ref{action}) also exhibits an 
${\rm SL}(2,R)$ `S--duality' \cite{wil}. In 
four dimensions, the three--form field strength of the NS--NS
two--form potential is dual to a one--form
corresponding to the field strength of a pseudo--scalar 
axion field, $\sigma$: 
\be
\label{Hsigma}
H^{\alpha\beta\gamma} \equiv e^{\Phi} 
\epsilon^{\alpha\beta\gamma\delta} \nabla_{\delta} 
\sigma    ,
\ee
where $\epsilon^{\alpha\beta\gamma\delta}$ is 
the covariantly constant antisymmetric 
four--form. Performing the conformal transformation 
\be
\label{conformal}
\tilde{g}_{\mu\nu} = \Theta^2 g_{\mu\nu}, 
\qquad \Theta^2 \equiv e^{-\Phi}
\ee
implies that the NS--NS action (\ref{action})
transforms to 
\be
\label{einact}
S= \int d^4x\sqrt{-\tilde{g}} \left[ 
\tilde{R} -\frac{1}{2} \left( \tilde{\nabla} \Phi 
\right)^2 -\frac{1}{2} e^{2\Phi} \left( \tilde{\nabla} \sigma 
\right)^2 \right]
\ee
and this may be written in the form of Eq. (\ref{actiontrace}) with 
$\Lambda =0$ by defining 
\be 
\label{NSL}
N \equiv  \left( \begin{array}{cc} e^{\Phi} & \sigma e^{\Phi} \\
\sigma e^{\Phi} & e^{-\Phi} +\sigma^2e^{\Phi}
\end{array} \right)    .
\ee
The dilaton and axion 
fields parametrize  the ${\rm SL}(2,R)/{\rm U}(1)$ 
coset. Since the inverse of the symmetric matrix (\ref{NSL}) is given by 
$N^{-1} = -JNJ$, where $J$ is defined in Eq. (\ref{J}), 
a supersymmetric extension of the FRW cosmologies may  also
be developed for this model. 

In conclusion, therefore, $N=2$ 
supersymmetric quantum cosmologies may be derived from the non--linear 
sigma--models associated with S--, T-- and U--dualities of  
string theory. 

\vspace{.3in}
\centerline{{\bf Acknowledgments}}
\vspace{.3in}

JEL is supported by the Particle Physics and 
Astronomy Research Council (PPARC) UK. JM would like to 
thank G. Veneziano for valuable discussions and encouragement 
and would like to acknowledge the gracious hospitality 
of the Theory Division of CERN.

\end{document}